\documentclass[11pt,reqno]{amsart}

\usepackage{amsfonts}
\usepackage{amssymb}
\usepackage{array}
\usepackage{xcolor}
\usepackage[colorlinks=true,linkcolor=blue!55!black,citecolor=blue!55!black,
            urlcolor=blue!55!black]{hyperref}

\newtheorem{theorem}{Theorem}[section]
\newtheorem{lemma}[theorem]{Lemma}
\newtheorem{proposition}[theorem]{Proposition}
\theoremstyle{definition}
\newtheorem{definition}[theorem]{Definition}
\theoremstyle{remark}

\newcommand{\Z}{\mathbb Z}

\numberwithin{equation}{section}

\newcommand{\abs}[1]{\lvert#1\rvert}
\newcommand{\cl}{\operatorname{cl}}

\newcommand{\Dh}{\mathrm{DH}}

\begin{document}
\allowdisplaybreaks
\abovedisplayskip=2pt plus 2pt minus 2pt
\belowdisplayskip=1pt plus 2pt minus 2pt
\abovedisplayshortskip=1pt plus 1pt
\belowdisplayshortskip=1pt plus 1pt

\title[A faster commutator-length algorithm]{A faster algorithm for commutator
length in free groups}

\author[Z.X. Wang]{Zixing Wang}

\thanks{Email address: \texttt{zixingwang.cn@gmail.com}}

\date{}



\begin{abstract}
We compute the commutator length $\cl(w)$ of a reduced word $w\in[F,F]$ of
length $M$ in a free group.  Given a target $g\ge1$, a deterministic algorithm decides
$\cl(w)\le g$ and, when it holds, returns a decomposition into at most $g$
commutators, in $O(M^{3g})$ steps; trying $g=1,2,\dots$ gives $\cl(w)$ in
$O(M^{3\cl(w)})$ steps.  For $g=1$ the same test takes $O(M)$ expected steps
on a uniformly random reduced input in $[F,F]$.  All bounds are in the unit-cost
model.
\end{abstract}

\maketitle

\section{Introduction}\label{sec:intro}

Let $F$ be the free group on a finite set $X$ with $\abs{X}\ge2$, the case of
one generator having $[F,F]=1$; throughout, $w$ is a reduced word over
$X\cup X^{-1}$ of length $M=\abs{w}$.  For $w\in[F,F]$ put
\begin{equation}\label{eq:cl}
\cl(w)=\min\Bigl\{k\ge0:\ w=\prod_{i=1}^{k}[u_i,v_i]\Bigr\},
\qquad [u,v]=uvu^{-1}v^{-1},
\end{equation}
and $\cl(w)=\infty$ when $w\notin[F,F]$, the set above being empty.
Throughout, $g\ge1$ is the \emph{target}, the bound on the number of
commutators given as part of the input: the task is to decide $\cl(w)\le g$
and, when that holds, to exhibit a decomposition.  The exponent below is linear in $g$, so for each fixed target the bounds are
polynomial in $M$, as in~\cite{BFI}.  At a node of the recursion $\mathit{rem}$ denotes the
target still available there.

Culler~\cite{Culler} identifies $\cl(w)$ with the least genus of a one-face
map with boundary $w$.  The best published bound we know is that of
Bartholdi, Fialkovski and Ivanov~\cite{BFI} (henceforth BFI): their
Theorem~A gives, for
non-trivial $w\in[F,F]$, a cancellation-free factorisation
\begin{equation}\label{eq:factor}
w=W_{1}a^{-1}W_{2}b^{-1}W_{3}aW_{4}bW_{5},
\quad a,b\in X\cup X^{-1},
\end{equation}
with $\cl(W_{1}W_{4}W_{3}W_{2}W_{5})=\cl(w)-1$.  Repeating those four cut
points at each of the $g$ levels costs $O(M^{4g})$ on an $F$-RAM (their
Corollary~C), and their {\S}2.4 improves the $\cl=1$ case to an average
$O(M^{2})$ by a $2\times2$ matrix trace test.

We keep this recursion and replace two of its ingredients.  \emph{Base case}:
one Manacher sweep builds an arc graph in which a word of class one shows up
as a walk of at most three arcs from a vertex to its antipode ($O(M^{3})$ worst
case, $O(M)$ in expectation, Theorem~\ref{thm:avg}).  \emph{Recursion}: a peel
is an exchange of two blocks, so three cut points replace four and the worst
case is $O(M^{3g})$.  That bound alone does not make the search complete:
Theorem~\ref{thm:peel} supplies, at every node, a peel lowering the class by
exactly one.

All bounds are in the \emph{unit-cost} model, one step per letter read, lookup
or comparison.  BFI count group operations, so a residual of length $M$ is $M$
steps here and one there, and their $O(M^{4g})$ is $O(M^{4g+1})$ steps: comparing
exponents alone is the conservative choice.

\begin{theorem}[main]\label{thm:main}
For a reduced $w\in[F,F]$ of length $M$ and a target $g\ge1$, a deterministic
algorithm returns a decomposition of $w$ into at most $g$ commutators, or
reports $\cl(w)>g$, in $O(M^{3g})$ steps and $O(M^{2})$ space.  The algorithm is
deterministic; for $g=1$ its expected running time is $O(M)$ when the input is
drawn uniformly from reduced words of length $M$ conditioned on lying in
$[F,F]$ --- the expectation is over inputs only.
\end{theorem}

For $w=1$ the answer is the empty decomposition; assume $w\ne1$ below, so
$M\ge4$.  Only targets $g\le\lfloor M/4\rfloor$ need be searched.  By
Culler~\cite{Culler} the class of $w$ is the genus $\kappa$ of a one-face map
whose edges are the inverse pairs of letters of $w$, so that map has
$E=M/2$ edges and Euler's formula gives $2\kappa=E-V+1\le E=M/2$; hence
$\cl(w)\le M/4$, and being an integer, $\cl(w)\le\lfloor M/4\rfloor$.  A target
$g>\lfloor M/4\rfloor$ is replaced by $\min(g,\lfloor M/4\rfloor)$ and searched
at that value: the search still returns a decomposition into at most
$\cl(w)\le\lfloor M/4\rfloor\le g$ commutators.  Throughout the search
$gM\le M^{2}$.

\section{The class-one case: an expected-linear base test}\label{sec:cl1}

\subsection{Notation and the cost model}\label{sec:pre}

Write $w=w_{0}\cdots w_{M-1}$ and read the indices mod $M$.  A \emph{c-word} is a word of class one.  By Wicks~\cite{Wicks} a cyclically
reduced word $w\ne1$ is a c-word iff some cyclic shift of it equals
$U\,V\,W\,U^{-1}V^{-1}W^{-1}$ without cancellation, $U,V,W$ being contiguous
blocks, possibly empty; in general $w$ is a c-word iff it is conjugate to such a
word.  The three cut points of that shift are what the base test looks for, on
the cyclically reduced residual.  A word of $[F,F]$ has even length and trivial abelianisation.
Since $\cl$ is conjugacy-invariant we first strip the longest
\emph{mirror-inverse prefix} $w[0\!:\!s]$, that is the largest $s=s_{\max}$ with
$w_{j}=w_{M-1-j}^{-1}$ for all $j<s$.  Then $w=x\,z\,x^{-1}$ with
$x=w[0\!:\!s_{\max}]$ and $z=w[s_{\max}\!:\!M-s_{\max}]$; for $w\in[F,F]$ reduced and non-trivial one has
$s_{\max}\le\mathit{half}-2$, where $\mathit{half}=M/2$, so $z$ keeps at least
four letters and is cyclically reduced (Lemma~\ref{lem:onepass}).

\emph{Cost model.}  Letters are integers.  An \emph{average} is taken over $w$
uniform among the $A(A-1)^{M-1}$ reduced words of length $M$ over
$X\cup X^{-1}$, where $A=2\abs{X}\ge4$; these are the words the algorithm is
handed, and the average of Theorem~\ref{thm:avg} is over those of them that lie
in $[F,F]$, that is, conditioned on $w\in[F,F]$.  Free reduction and the test
$w\in[F,F]$ are $O(M)$ stack passes, dominated by every bound below.

\emph{Pruning invariant.}  Let $r$ be a reduced residual and write
$r=x\,z\,x^{-1}$, where $z$ is its cyclically reduced core, unique up to cyclic
rotation.  For an ordered pair $U=ab$ of letters, let $b_{U}$ be the number of
cyclic positions $q$ of $z$ with $z_{q}z_{q+1}=U$, indices read mod $\abs{z}$.
No such block is its own inverse, since $z$ is cyclically reduced.  The
\emph{defect} is
\begin{equation}\label{eq:count}
\Dh(r)=\sum_{\{U,U^{-1}\}}\abs{b_{U}-b_{U^{-1}}}.
\end{equation}
A cyclic rotation of $z$ permutes the counted blocks, so $\Dh(r)$ does not
depend on the chosen rotation of the core or on where $r$ was anchored.  We use
the following self-contained bound.

\begin{lemma}[defect bound]\label{lem:defect}
For every reduced non-trivial $r\in[F,F]$,
\begin{equation}\label{eq:D}
\cl(r)\ge\frac{\Dh(r)+6}{12}=\frac{\Dh(r)}{12}+\frac12 .
\end{equation}
\end{lemma}

\begin{proof}
Write $r=xzx^{-1}$ as above, $m=\abs{z}=2n$, and let $\kappa=\cl(z)=\cl(r)$.  Choose
a pairing $\pi$ attaining the maximum in Proposition~\ref{prop:cycles} below
and set $\sigma=\rho\pi$, with $\rho=(0\,1\,\cdots\,m-1)$ and
$c=c(\sigma)$; then $2\kappa=n+1-c$.  Cyclically reducedness gives $\sigma(q)\ne q$
for all $q$, since $\sigma(q)=q$ would mean $\pi(q)=q-1$, that is, adjacent
inverse letters across the cyclic seam of $z$.

A block $z_{q}z_{q+1}$ has an inverse block at $\pi(q+1)$ precisely when
$\pi(q)=\pi(q+1)+1$, equivalently $\sigma^{2}(q+1)=q+1$; as $\sigma$ has no
fixed points, this happens exactly when $q+1$ lies in a $2$-orbit of $\sigma$.
Each $2$-orbit $\{a,b\}$ yields exactly the two matched block positions
$a-1,b-1$, one block and its inverse.  Let $t$ be the number of $2$-orbits, so
$2t$ positions are matched.  Fix an inverse-pair class $\{U,U^{-1}\}$ with
original counts $b,b'$ and $c_{U}$ matched blocks of each kind.  Removing the
matched pairs subtracts the same $c_{U}$ from both counts, so the imbalance stays
$\abs{b-b'}$; the unmatched positions of the class number
$(b-c_{U})+(b'-c_{U})=\abs{b-b'}+2(\min(b,b')-c_{U})\ge\abs{b-b'}$.  Summing the imbalance
over classes gives at most the total number $m-2t$ of unmatched positions:
$\Dh(z)\le m-2t$.

Let the remaining orbits have lengths $L_{1},\dots,L_{s}\ge3$.  Then
$m=2t+\sum_i L_i$ and $m-2c=\sum_i(L_i-2)$.  Since $L_i\le3(L_i-2)$ for
$L_i\ge3$,
\[
\Dh(r)\le\sum_i L_i\le3\sum_i(L_i-2)=3(m-2c)=12\kappa-6,
\]
which is \eqref{eq:D}.
\end{proof}

In general, a residual whose \emph{own} target is $t$ is rejected at
$\Dh\ge12\,t-5$: then \eqref{eq:D} gives $\cl\ge t+\tfrac1{12}$, hence
$\cl\ge t+1$ above its target.  Thus at the root $t=\mathit{rem}$, while a
child of a node with target $\mathit{rem}$ has $t=\mathit{rem}-1$.

\subsection{The arc graph and a three-hop walk}\label{sec:base}

\begin{definition}\label{def:arc}
Let $M$ be even, $\mathit{half}=M/2$.  There is an \emph{arc of length
$\ell$} from $x$, written $x\to x+\ell$ ($1\le\ell\le\mathit{half}$), if
\[
w_{x+t}=w_{x+\mathit{half}+\ell-1-t}^{-1}\qquad(0\le t<\ell),
\]
indices mod $M$.  The \emph{arc graph} $G(w)$ has vertex set $\Z/M$, with an
arc $x\to x+\ell$ for every pair $(x,\ell)$ satisfying the condition.
\end{definition}

On
$abc\,a^{-1}b^{-1}c^{-1}$ three unit arcs give $0\to1\to2\to3$, while $[ab,c]$
needs the length-$2$ arc $0\to2$.  For fixed $x$ the conditions are not nested in
$\ell$, but for a fixed \emph{centre} they are: the arc $x\to x+\ell$ has centre
\begin{equation}\label{eq:centre}
c\equiv2x+\ell-1\pmod M,
\end{equation}
and its $\ell$ conditions are one palindromic sweep of an interleaved word about
$c+\tfrac12$.

\begin{lemma}[the arcs of one centre]\label{lem:centre}
Read $v_{2j}=w_{j}$, $v_{2j+1}=w_{\mathit{half}+j}^{-1}$ on $\Z/2M$, so that $v$
has length $2M$.  The arc
$x\to x+\ell$ of centre $c$ exists iff the $2\ell$ positions
$2x,\dots,2x+2\ell-1$ of $v$ form a palindrome of even radius $\ell$ about
$c+\tfrac12$; hence, $r_{c}$ being the radius of $v$ there, the arcs of centre
$c$ are the lengths $\ell\le\min(r_{c},\mathit{half})$ with
$\ell\equiv c+1\pmod2$.
\end{lemma}

\begin{proof}
With $q=x+\mathit{half}+\ell-1-t$ the condition is $w_{x+t}=w_{q}^{-1}$, that
is $v_{2x+2t}=v_{2(q-\mathit{half})+1}$: by $v_{2j+1}=w_{\mathit{half}+j}^{-1}$
the letter $w_{q}^{-1}$ sits at $v_{2(q-\mathit{half})+1}$, and
$2(q-\mathit{half})+1=2x+2\ell-1-2t$.  The two indices add up to $2c+1$: each
condition is one mirror pair about $c+\tfrac12$, and the $\ell$ of them walk
through the block $[2x,2x+2\ell-1]$ from its ends inwards.  Conversely such a
palindrome meets all $\ell$ conditions of the arc at $x\equiv(c+1-\ell)/2$,
so $x$ is an integer exactly when $\ell\equiv c+1\pmod2$.

The centre of \eqref{eq:centre} is a class modulo $M$; in $v$, whose indices
are modulo $2M$, it has the two lifts $c+\tfrac12$ and $c+M+\tfrac12$.
That is what identifies the two twin sources: the arc at $x$ has its
palindrome centre at $c+\tfrac12$ in $v$, the arc at $x+\mathit{half}$ at the antipodal
$c+M+\tfrac12$, and the two radii agree on lengths
$\ell\le\mathit{half}$, because the shift $x\mapsto x+\mathit{half}$ carries
arcs to arcs and the mirror pairs about $c+M+\tfrac12$ read
$w_{y}=w_{\mathit{half}+c-y}^{-1}$ as well --- the same $\ell$ conditions, read
at the antipodal positions.  The full radii of $v$ at the two centres need not
agree, but only lengths up to $\mathit{half}$ are arcs, so one radius per
centre suffices and the sweep of $v$ sees every arc.
\end{proof}

So a length determines its two sources, $x$ and $x+\mathit{half}$.  One Manacher sweep~\cite{Manacher} on three copies of
$v$, a word of length $6M$, so that no radius is cut off at an end, computes all
circular radii in $O(M)$ and lists the arcs of a centre in time proportional to their number: the
two sources are twins sharing the centre, and an arc's centre is determined by
its source and its length, so nothing is emitted twice.
Building $G$ costs $O(M+\nu)$ with $\nu=\abs{E(G)}$; let $\Delta$ be the
maximum degree, in or out, and $Q=\sum_{v}\deg^{-}(v)\deg^{+}(v)$ the number of
two-arc paths, so $Q\le\Delta\nu$.

\emph{The search.}  The shift $x\mapsto x+\mathit{half}$ carries the arc
$(x,\ell)$ to $(x+\mathit{half},\ell)$, so it is an automorphism of $G$ and it
suffices to test the $\mathit{half}$ starts $i<\mathit{half}$.  Arcs are read
modulo $M$, so the intermediate vertices of a longer walk may wrap; in this
range the only arc that could be missed is the single arc
$i\to i+\mathit{half}$ of length $\mathit{half}$, which cyclic reducedness
forbids as in the proof of Proposition~\ref{prop:3hop}.  So only two and three
hops need testing.  Mark the in-neighbours
of $i+\mathit{half}$ in a Boolean array that carries a version stamp, so that a
slot counts as marked only during the start that wrote it and no array is
cleared between starts; then run through the arcs out of $i$.  A first hop
$i\to u$ is already a hit if $u$ is marked; otherwise the arcs out of $u$ are
looked up in the array one by one, $\deg^{+}(u)$ queries of $O(1)$.  Every walk of at most three arcs
from $i$ is inspected this way, and the first hit returns its intermediate
vertices as cut points.

\begin{proposition}[Wicks' criterion as a walk]\label{prop:3hop}
Let $w\ne1$ be a cyclically reduced word of even length.  Then $\cl(w)=1$
iff, for some $i<\mathit{half}$, $G(w)$ has a walk of at most three arcs from
$i$ to $i+\mathit{half}$; the intermediate vertices are the cut points.
\end{proposition}

\begin{proof}
If $w=U\,V\,W\,U^{-1}V^{-1}W^{-1}$ without cancellation, with
$p_{1}=0$, $p_{2}=\abs{U}$, $p_{3}=\abs{UV}$, the three blocks give arcs
$p_{1}\to p_{2}$, $p_{2}\to p_{3}$, $p_{3}\to p_{1}+\mathit{half}$ of lengths
$\abs{U},\abs{V},\abs{W}$, which sum to $\mathit{half}$; an empty block simply
drops an arc.

Conversely, let a walk of at most three arcs lead from $i$ to $i+\mathit{half}$,
and pad it by repeating vertices until it reads $i\to y\to z\to
i+\mathit{half}$, with lengths $\ell_{1},\ell_{2},\ell_{3}\ge0$; a length $0$
is the degenerate arc at a repeated vertex and contributes an empty block.
Their sum lies in $[0,3\mathit{half}]$ and is
$\equiv\mathit{half}\pmod M$, hence equals $\mathit{half}$ or $3\mathit{half}$.
No arc of length $\mathit{half}$ can occur: such an arc out of any $x$ requires
$w_{x+\mathit{half}-1}=w_{x+\mathit{half}}^{-1}$, and those two letters are
cyclically adjacent in a cyclically reduced $w$.  Hence the sum cannot be
$3\mathit{half}$, and when it is $\mathit{half}$ every $\ell_{i}<\mathit{half}$:
the single-arc case $(\mathit{half},0,0)$ and its permutations are excluded
too, and the walk is monotone.

Now cut at $y$ and $z$.  We get $U=w[i{:}y]$,
$V=w[y{:}z]$, $W=w[z{:}i+\mathit{half}]$, and the arc conditions say that the
antipodal half reads $U^{-1}V^{-1}W^{-1}$.  Every junction but the base one
joins adjacent letters of the reduced word $w$; the base one does not cancel,
again by cyclic reducedness.  That is Wicks' form, so $\cl(w)=1$.  The
decomposition is explicit: $UV\,W\,U^{-1}V^{-1}W^{-1}=[UV,\,WU^{-1}]$, so the
search returns $u=UV$ and $v=WU^{-1}$.  These decompose the cyclically reduced
residual; a node holding $r=xzx^{-1}$ returns $[xux^{-1},xvx^{-1}]$.
\end{proof}

\begin{proposition}[correctness and worst case]\label{prop:base}
Let $w$ be as in Proposition~\ref{prop:3hop}.  If $w$ is a c-word, the build
and search above return its cut points.  One pass costs $O(M+\nu+Q)$, hence
$O(M^{3})$ in the worst case.
\end{proposition}

\begin{proof}
The build is $O(M+\nu)$ and correctness is Proposition~\ref{prop:3hop}.  For the search, a start $i$ pays
$\deg^{-}(i+\mathit{half})$ for the marking, and $\deg^{+}(u)$ for each arc
$i\to u$ out of it; summing over the starts gives at most $\nu+Q$ per pass, and
the build adds $\nu$.  With $\nu\le M\,\mathit{half}\le M^{2}$ (one arc of
each length per vertex) and $Q\le\Delta\nu\le M^{3}$, that is $O(M^{3})$.

The bound is attained on $[a,b^{m}]^{3}$, $M=6m+6$, which is not a $c$-word, so
every one of the $\mathit{half}$ starts is examined.  Each of its three $b^{m}$
runs faces a $b^{-m}$ run at the antipodal position, and for each length
$2\le\ell\le m$ this yields $6(m-\ell+1)$ arcs of length $\ell$, plus $M$ arcs of
length one; hence $\nu=M+3m(m-1)=3m^{2}+3m+6=\Theta(M^{2})$.  Convolving the
in/out degrees of this arc family gives $Q=m^{3}+11m=\Theta(M^{3})$.
\end{proof}

\subsection{The expected cost}\label{sec:avg}

\begin{lemma}[the longest strip suffices]\label{lem:onepass}
Let $w\in[F,F]$ be reduced, non-trivial and of even length.  Then
$s_{\max}\le\mathit{half}-2$, the residual $z=w[s_{\max}{:}M-s_{\max}]$ is
cyclically reduced, and $\cl(z)=\cl(w)$; so the base test decides $\cl(w)=1$
correctly.
\end{lemma}

\begin{proof}
Put $M=2s+m$ with $s=s_{\max}$ and $z=w[s{:}M-s]$, so $\abs{z}=m$ and
$w=xzx^{-1}$ with $x=w[0{:}s]$; conjugation gives $\cl(z)=\cl(w)$.  The
mirror condition holds at level $j<s$ exactly when $w_{j}=w_{M-1-j}^{-1}$.
At the next level $j=s$ it reads $w_{s}=w_{M-1-s}^{-1}$, and since
$z_{0}=w_{s}$ and $z_{m-1}=w_{M-s-1}=w_{M-1-s}$, this is precisely
$z_{0}=z_{m-1}^{-1}$: the two ends of the residual are inverse.  Thus if $z$
were not cyclically reduced the prefix would extend to $s+1$, contradicting
maximality.  This covers $s=0$, where $z=w$, and $m=2$, where a reduced
two-letter word is automatically cyclically reduced.

If $s\ge\mathit{half}$, then $j=\mathit{half}-1<s$ and the condition at that
level reads $w_{\mathit{half}-1}=w_{M-\mathit{half}}^{-1}
=w_{\mathit{half}}^{-1}$, which reducedness forbids; hence
$s_{\max}\le\mathit{half}-1$.  Equality would give $m=2$ and
$z=w_{\mathit{half}-1}w_{\mathit{half}}$, which is reduced, hence not
$aa^{-1}$, and has non-trivial abelianisation, contradicting $w\in[F,F]$.
Therefore $s\le\mathit{half}-2$, $z$ is cyclically reduced, and
Proposition~\ref{prop:3hop} decides $\cl(z)=1$, that is $\cl(w)=1$.
\end{proof}

\begin{lemma}[the abelianisation of a random reduced word]\label{lem:cond}
Let $z$ be uniform among the reduced words of length $m$ over $X\cup X^{-1}$,
and let $S(z)=\sum_{i}z_{i}\in\Z^{\abs{X}}$ be its abelianisation, so that
$z\in[F,F]$ reads $S(z)=0$.  Then
(i) the letters form a Markov chain, $z_{0}$ uniform on $X\cup X^{-1}$ and,
given $z_{i-1}$, $z_{i}$ uniform among the $A-1$ letters not inverse to
$z_{i-1}$;
(ii) there are $c_{1},c_{2},c_{3}>0$, depending only on $X$, with
\[
c_{1}m^{-\abs{X}/2}\le\mathbb P(S(z)=\delta)\le c_{2}m^{-\abs{X}/2}
\qquad\text{if }\lVert\delta\rVert\le c_{3}\sqrt m
\]
for every $\delta$ whose parity is compatible with $m$, and the upper bound
$\mathbb P(S(z)=\delta)\le c_{2}m^{-\abs{X}/2}$ holds for all $\delta$;
(iii) if $B$ is a union of at most $K$ intervals of positions with
$\abs{B}\le m/2$, and $u$ is a pattern on $B$, then
\[
\mathbb P(z|_{B}=u\mid S(z)=0)\le c_{4}\,\mathbb P(z|_{B}=u),
\]
where $c_{4}$ depends only on $X$ and $K$; moreover $c_{2}/c_{1}\to1$ as
$m\to\infty$.
\end{lemma}

\begin{proof}
(i) is the measure itself: $z_{0}$ is uniform on $X\cup X^{-1}$, and each
subsequent letter is uniform on the $A-1$ letters other than the inverse of
its predecessor.

\emph{Proof of (ii).}  Observe first that $S(z)=\sum_{i}z_{i}$ is an additive
functional of the aperiodic irreducible chain of (i).  Its increment has
stationary mean zero and stationary increment covariance $\frac{1}{\abs{X}}I$;
the transition operator has eigenvalue $-1/(2\abs{X}-1)$ on the mean-zero
symmetric subspace and $+1/(2\abs{X}-1)$ on the antisymmetric one, so the chain
is uniformly ergodic.  The observable $z_{i}\in\Z^{\abs{X}}$ is antisymmetric,
hence lies in the latter eigenspace, with $\lambda=1/(2\abs{X}-1)$, and
$\mathrm{Var}(S(z))\sim\frac{m}{\abs{X}-1}I=\frac{1}{\abs{X}}\frac{1+\lambda}{1-\lambda}mI\succ0$.

Fourier inversion now gives $\mathbb P(S(z)=\delta)\asymp m^{-\abs{X}/2}$
uniformly for $\lVert\delta\rVert=O(\sqrt m)$.  Write
$\varphi(\theta)=\mathbb E[e^{\mathrm{i}\langle\theta,S(z)\rangle}]$.  Since
every letter contributes $\pm e_{x}$, $S(z)$ lies in the index-two coset
$L_{m}=\{\delta:\sum_{x}\delta_{x}\equiv m\pmod2\}$.  Let $\mathrm{ab}(d)=\pm e_{x}$
denote the vector of a letter $d$, so that
$\hat P_{\theta}(c,d)=P(c,d)e^{\mathrm{i}\langle\theta,\mathrm{ab}(d)\rangle}$
and $\abs{\hat P_{\theta}}=P$.  Since $P$ is irreducible and aperiodic,
Wielandt's equality case forces $\hat P_{\theta}=\lambda DPD^{-1}$, with
$\abs{\lambda}=1$ and $D$ diagonal unitary, whenever $\rho(\hat P_{\theta})=1$.
Entrywise this reads
$e^{\mathrm{i}\langle\theta,\mathrm{ab}(d)\rangle}=\lambda e^{\mathrm{i}(\phi_{c}-\phi_{d})}$
for every $d\ne c^{-1}$; the left side is independent of $c$ and each $d$ has
$A-1\ge3$ predecessors, so all $\phi_{c}$ coincide and
$e^{\mathrm{i}\langle\theta,\mathrm{ab}(d)\rangle}=\lambda$ for every $d$.
Taking $d=x$ and $d=x^{-1}$ gives
$e^{\mathrm{i}\theta_{x}}=e^{-\mathrm{i}\theta_{x}}=\lambda$: all $\theta_{x}$
share one common value $c=\pm1$, so the only characters of modulus one are
$\theta=0$ and $\theta_{\ast}=(\pi,\dots,\pi)$.  Near $\theta=0$,
$\abs{\varphi(\theta)}\le\exp(-cm\lVert\theta\rVert^{2})$, and $\theta_{\ast}$
carries the same peak since $\varphi(\theta_{\ast}+\epsilon)=(-1)^{m}\varphi(\epsilon)$;
the two add on $L_{m}$ and cancel off it.  Away from both,
$\abs{\varphi(\theta)}\le\varrho^{m}$ ($\varrho<1$).  The main term is the same
Gaussian integral in the upper and lower bounds, whence $c_{2}/c_{1}\to1$; the
same bounds give the upper estimate for every $\delta$, with no restriction on
its size.

\emph{Proof of (iii).}  \enlargethispage{5\baselineskip}
Write the complement of $B$ as $K+1$ gaps, some empty,
with $r_{1},\dots,r_{K+1}\ge0$ free letters and
$r=\sum_j r_j=m-\abs{B}\ge m/2$.  An interior gap has both flanking letters
fixed by $u$; only the two end gaps have an endpoint outside $B$.  Condition
further on a complete assignment $\beta$ of those at most two outer endpoint
letters.  Given $u$ and $\beta$, the letters in distinct gaps are independent:
a gap with $r_j\ge1$ free letters is a Markov bridge of $r_j+1$ transitions
between its fixed endpoints, its increment weight being the corresponding
product of the refined transition matrix $P(\cdot,\cdot;s)$,
$s\in\Z^{\abs{X}}$ (the fixed right endpoint shifts $s$ by a constant); an
empty gap contributes the point mass $1$ at $0$.

The spectral gap gives $P^{t}(a,b)=\pi(b)(1+O(\varrho^{t}))$ uniformly in
$a,b$, so fixed endpoints cost a factor bounded independently of them; the
Fourier estimate of (ii) applies to the refined product and gives
$\mathbb P(S_{\mathrm{gap},j}=s\mid u,\beta)\le c_{2}r_j^{-\abs{X}/2}$
uniformly in $s$ and in the endpoints when $r_j\ge1$.  Since
$S(z)=S_{B}(u)+\sum_j S_{\mathrm{gap},j}$ and the gap increments are
independent given $u,\beta$, the conditional law of $S(z)$ is the convolution
of the $K+1$ gap laws, shifted by $S_{B}(u)$.  A convolution's point mass is at
most the largest point mass of any factor, and some gap has
$r_j\ge r/(K+1)\ge m/(2K+2)$, giving
$\mathbb P(S(z)=0\mid u,\beta)\le c_{5}m^{-\abs{X}/2}$ uniformly in $\beta$.

Averaging over the at most $A^{2}$ values of $\beta$ preserves this bound, so
$\mathbb P(S(z)=0\mid z|_{B}=u)\le c_{5}m^{-\abs{X}/2}$.  Dividing by the lower
bound of (ii) at $\delta=0$ (here $m$ is even) proves (iii).
\end{proof}

\begin{theorem}[expected linear time]\label{thm:avg}
The $\cl=1$ test runs in $O(M)$ expected steps on a uniformly random reduced
word of length $M$, conditioned on $w\in[F,F]$, in the unit-cost model.
\end{theorem}

\begin{proof}
There is a constant $C>0$ with
\begin{equation}\label{eq:det}
T(w)\le C\,(M+\nu+Q)
\end{equation}
for every $w$: the sweep is $O(M)$, emitting the arcs is $O(\nu)$, and the
walk costs at most $\nu+Q$ by Proposition~\ref{prop:base}.

\emph{The unconditional measure.}  We first estimate in the uniform measure on
reduced residuals $z$ of length $m$, so $\mathit{half}=m/2$ here; at the end
$m\le M$.  Write $x:=1/(A-1)$.  For a fixed arc
candidate $(x_{0},\ell)$ the $\ell$ conditions pair the positions $x_{0}+t$ and
$x_{0}+\mathit{half}+\ell-1-t$ for $0\le t<\ell$.  These pairs are disjoint,
and a pair whose positions are adjacent has probability $0$ by reducedness.
Otherwise reveal the word from left to right: when the later endpoint of a
pair is reached, the earlier letter is already known, and the Markov rule
prescribes its inverse with probability either $0$ (if forbidden by the
preceding letter) or exactly $x$.  Multiplying over the $\ell$ pairs gives
$\mathbb P(\text{arc of length }\ell)\le x^{\ell}$.

A vertex carries at most one arc of each length, so a union bound over the
$m\,\mathit{half}$ candidate arcs gives
$\mathbb E[\nu]\le m\sum_{\ell\ge1}x^{\ell}=mx/(1-x)$.  A two-arc path with
lengths $(\ell_{1},\ell_{2})$ presupposes the longer of its two arcs, hence has
probability at most $x^{\max(\ell_{1},\ell_{2})}$, so
\[
\mathbb E[Q]\le m\!\!\sum_{\ell_{1},\ell_{2}\ge1}\!\!x^{\max(\ell_{1},\ell_{2})}
=m\,\frac{x(1+x)}{(1-x)^{2}} .
\]
Adding the three terms of \eqref{eq:det} applied to $z$ gives
$\mathbb E[T]\le C\bigl(1+\frac{x}{1-x}+\frac{x(1+x)}{(1-x)^{2}}\bigr)m<3Cm\le3CM$
for $A\ge4$: a bound linear in $M$.

\emph{Two conditionings.}  The measure of the theorem differs from the one the
test actually runs under in two ways; we handle them in turn.  First fix
$k=s_{\max}$.  Writing $w=x\,z\,x^{-1}$ with $\abs{x}=k$
and $\abs{z}=m=M-2k$, we have $s_{\max}=k$ exactly when $z_{0}\ne z_{m-1}^{-1}$
and $x$ is reduced with $x_{k-1}\notin\{z_{0}^{-1},z_{m-1}\}$, the two junctions.
Indeed $x^{-1}$ reads $x_{k-1}^{-1}\cdots x_{0}^{-1}$, so
$w_{M-1-j}=x_{j}^{-1}=w_{j}^{-1}$ for every $j<k$ whatever $x$ is; the prefix
extends to $j=k$ only when $z_{0}=z_{m-1}^{-1}$.  Under $z_{0}\ne z_{m-1}^{-1}$
the two forbidden last letters of $x$ are distinct.  For $k\ge1$ a reduced word
of length $k$ has $(A-1)^{k-1}$ continuations for each prescribed last letter,
so exactly $(A-2)(A-1)^{k-1}$ words $x$ are admissible; for $k=0$ the only
admissible $x$ is the empty word, a count of one.  Distinct $x$ give distinct
$w$.  For each fixed $k$ the tilt is therefore a constant, independent of $z$:
conditional on $s_{\max}=k$, $z$ is uniform among the reduced words of length
$m$ with $z_{0}\ne z_{m-1}^{-1}$.  The event
$E=\{z_{0}\ne z_{m-1}^{-1}\}$ has probability bounded away from $0$ and $1$:
the letter chain has $\mathbb P(z_{0}=a,z_{m-1}=b)=(1/A)P^{m-1}(a,b)=(1+o(1))/A^{2}$,
so $\mathbb P(E)=1-A^{-1}+o(1)$.

\emph{The abelianisation conditioning.}  Second, as $x$ and $x^{-1}$ cancel,
$w\in[F,F]$ reads $S(z)=0$.  Intervals are
those of the linear word $z$, of length $m$; a block wrapping across the end
splits into at most two, so an arc occupies at most four intervals and a
two-arc path at most eight.  We need the conditioning on $S(z)=0$ only through
its denominator.  Write $N_{t}=\#\{z\in\mathcal R_{t}:S(z)=0\}$, where
$\mathcal R_{t}$ is the set of reduced words of length $t$.  A word with
$z_{0}=z_{m-1}^{-1}$ is $z=a\,y\,a^{-1}$ with $y\in\mathcal R_{m-2}$ and
$S(y)=0$; as $a$ has $A$ choices this gives
$\mathbb P(E^{c}\mid S(z)=0)\le A\,N_{m-2}/N_{m}$.  By (ii), for even
$t\ge t_{0}$ one has $N_{t}\asymp A(A-1)^{t-1}t^{-\abs{X}/2}$, with $t_{0}$
depending only on $X$, not on $m$ or $M$; only even $t$ can have $N_{t}>0$.  For
the finitely many even $m<t_{0}$ the cost is bounded by a constant depending
only on $X$, absorbed by enlarging $C$ in \eqref{eq:det}.  Hence, for $m\ge t_{0}$,
\begin{align*}
\mathbb P(E^{c}\mid S(z)=0)
&\le\frac{A}{(A-1)^{2}}\frac{c_{2}}{c_{1}}
   \Bigl(\frac{m}{m-2}\Bigr)^{\abs{X}/2}  \\
&=\frac{A}{(A-1)^{2}}+o(1)\le\frac49+o(1)<\frac12 .
\end{align*}
Hence $\mathbb P(E\mid S(z)=0)\ge\tfrac12$, and Bayes with
$\mathbb P(S(z)=0)\asymp m^{-\abs{X}/2}$ from (ii) gives
$\mathbb P(S(z)=0\mid E)\asymp m^{-\abs{X}/2}$ uniformly in $k$.

\emph{Transferring the pattern estimate.}  Let $A=\{z|_{B}=u\}$ be a pattern
event with $\abs{B}\le m/2$.  Bayes and $\mathbb P(E\mid S(z)=0)\ge\tfrac12$
give
\[
\mathbb P(A\mid E,S(z)=0)
=\frac{\mathbb P(A\cap E\mid S(z)=0)}{\mathbb P(E\mid S(z)=0)}
\le\frac{\mathbb P(A\mid S(z)=0)}{\mathbb P(E\mid S(z)=0)}
\le2c\,\mathbb P(A),
\]
using (iii) in the form $\mathbb P(A\mid S(z)=0)\le c\,\mathbb P(A)$; the bound
is relative to the \emph{unconditional} pattern probability (not to
$\mathbb P(A\mid E)$, which need not hold).  With $K=8$, a pattern on a union
$B$ of at most eight intervals with $\abs{B}\le m/2$ therefore obeys
$\mathbb P(z|_{B}=u\mid E,S(z)=0)\le c'\,\mathbb P(z|_{B}=u)$ with $c'=2c$: the
$m^{-\abs{X}/2}$ from the bridge convolution cancels the same factor in the
denominator.  An arc or a path is specified by its position set, a union of at
most eight intervals (for a path $\abs{B}\le2(\ell_{1}+\ell_{2})$).  The
patterns $u$ compatible with a fixed position set define disjoint events, and
$c'$ does not depend on $u$: fix the position set $B$ of a candidate arc or path
first, then sum the display over all $u$ compatible with $B$, which costs at
most the same constant since $\sum_{u}\mathbb P(z|_{B}=u)\le1$.  Hence the arc
bound survives as soon as $\ell\le m/4$ and the path bound as soon as
$\ell_{1}+\ell_{2}\le m/4$.

The remaining patterns have $\abs{B}>m/2$: an arc then has $\ell>m/4$, and a
two-arc path has $\max(\ell_{1},\ell_{2})>m/8$.  They are counted on the
residual, of length $m$, so there are $O(m^{2})$ arcs and $O(m^{3})$ paths;
each has probability at most $x^{\ell}$ or $x^{\max(\ell_{1},\ell_{2})}$, and
conditioning multiplies this by at most
$\bigl(\mathbb P(E)\,\mathbb P(S(z)=0\mid E)\bigr)^{-1}=O(m^{\abs{X}/2})$: the
Bayes factor over $E$ costs $\mathbb P(E)^{-1}=O(1)$, since $\mathbb P(E)$ is
bounded below, and the denominator $\mathbb P(S(z)=0\mid E)$ was just computed.  Since
$m^{O(1)}x^{m/8}$ is uniformly bounded in $m$, their total contribution is
$O(1)\le O(M)$.  Thus the expectation in \eqref{eq:det} is $O(M)$ in the
conditional measure too, with larger constants; here $m=M-2k\le M$, and the
strip and the full-input sweep are already covered by \eqref{eq:det}.
\end{proof}

\section{Higher class: recursion and completeness}\label{sec:prune}

\subsection{Peels are exchanges}\label{sec:peel}

A \emph{peel} is a triple of cut points $0\le i\le j\le k\le M$: it splits
$w=A\,B\,C\,D$ into four blocks and keeps the reduced word
\begin{equation}\label{eq:peel}
z=\mathrm{red}(A\,C\,B\,D)
\end{equation}
as the residual.  There are $\binom{M+3}{3}=\Theta(M^{3})$ triples, one nested
loop fewer than the four cut points of \eqref{eq:factor} require; the range is
the whole one, degenerate triples with an empty block included, and the exact
peel of Theorem~\ref{thm:peel} is among them.

\begin{proposition}[a peel removes one commutator]\label{prop:peel}
With $u=ABA^{-1}$, $v=ACA^{-1}$,
\begin{equation}\label{eq:exch}
w=[u,v]\,z,\qquad z=\mathrm{red}(A\,C\,B\,D),
\end{equation}
so $\cl(w)\le\cl(z)+1$.  Free reduction never lengthens a word, so
$\abs{z}\le\abs{w}$, with strict inequality as soon as one of the junctions
$A|C$, $C|B$, $B|D$ cancels.
\end{proposition}

\begin{proof}
Multiplying out and cancelling gives, in the free group,
\[
[u,v]\,A\,C\,B\,D=A\,B\,C\,B^{-1}C^{-1}C\,B\,D=A\,B\,C\,D=w;
\]
these equalities survive reduction.

For the length bound, before reduction $A\,C\,B\,D$ has the
length of $w$; reduction never increases it, and a cancelling junction makes
it strictly smaller (a cascade may delete more than two letters).  This proves
the proposition.
\end{proof}

\subsection{Exact peels and completeness}\label{sec:exact}

Proposition~\ref{prop:peel} gives only $\cl(w)\le\cl(z)+1$, while the
recursion needs a peel that lowers the class by exactly one at every node.

The proof runs in four steps, one per lemma below: pairings give the orbit
count that computes $\cl$ (Proposition~\ref{prop:cycles}); a peel
left-multiplies by a $3$-cycle (Lemma~\ref{lem:three}); the gain is two exactly
for one cyclic order of its entries (Lemma~\ref{lem:tau}); a descent count
produces such entries (Lemma~\ref{lem:exists}).

\begin{proposition}[the cycle-count formula, after Culler and Bardakov]%
\label{prop:cycles}
Let $w=w_{0}\cdots w_{M-1}\ne1$ be reduced with $w\in[F,F]$ and $M=2n\ge2$, let
$\rho=(0\,1\,\cdots\,M-1)$.  Call a fixed-point-free involution $\pi$ of
$\{0,\dots,M-1\}$ with $w_{\pi(p)}=w_{p}^{-1}$ a \emph{pairing}.  Then
\[
\cl(w)=\tfrac12\bigl(n+1-\max_{\pi}c(\rho\pi)\bigr),
\]
the maximum over all pairings, where $c$ counts orbits.
\end{proposition}

\begin{proof}
Pairings exist because the abelianisation of $w$ vanishes.
Place one dart for each letter of $w$ around a single vertex, in the order
$\rho$, and glue them in pairs as $\pi$ prescribes.  The resulting map has one
vertex and $n$ edges.  Walking a face alternates crossing an edge, which is
$\pi$, with turning to the next dart, which is $\rho$; so the faces are the
orbits of $\rho\pi$ and $F=c(\rho\pi)$.  Euler's formula $V-E+F=2-2\gamma$ with
$V=1$ and $E=n$ gives
$\gamma=\tfrac12(n+1-c(\rho\pi))$, and Culler's theorem, in the permutation form
given by Bardakov~\cite{Bardakov}, identifies $\cl(w)$ with the least genus of
such a map.  Minimising the genus means maximising $c(\rho\pi)$.
\end{proof}

Let $\pi$ be an optimal pairing, one maximising $c(\rho\pi)$, and put
\[
\sigma:=\rho\pi,\qquad \kappa:=\cl(w),\qquad c_{0}:=c(\sigma)=n+1-2\kappa .
\]
Since $\cl$ is a conjugacy invariant we may take $w$ cyclically reduced: if
$w=u\,w'\,u^{-1}$ and $w'=A'B'C'D'$ has $\cl(A'C'B'D')=\kappa-1$, then
$A=uA'$, $B=B'$, $C=C'$, $D=D'u^{-1}$ give $ACBD=u(A'C'B'D')u^{-1}$, of the
same class.  A pairing pairs inverse letters, so cyclically reducedness
restricts $\pi$ twice: $\pi(p)=p+1$ would force $w_{p+1}=w_{p}^{-1}$ at adjacent
positions, and $\pi(0)=M-1$ would force $w_{M-1}=w_{0}^{-1}$ across the cyclic
seam; both are forbidden.  Hence every pair
$\{p,q\}$ with $p<q$ has $q\ge p+2$ and $\{p,q\}\ne\{0,M-1\}$, and since
$\sigma(p)=\pi(p)+1\pmod M$, the permutation $\sigma$ has no fixed point.

\begin{lemma}[a peel multiplies by a $3$-cycle]\label{lem:three}
Let $w=A\,B\,C\,D$ with cut points $0\le i<j<k\le M-1$, let
$z_{0}=A\,C\,B\,D$, and let $\pi$ be a pairing of $w$.  Transporting $\pi$ along
the block move $\phi\colon A\,B\,C\,D\mapsto A\,C\,B\,D$ gives a map on $z_{0}$
whose face permutation has the orbit count of $\tilde\rho\pi$, where
$\tilde\rho:=\phi^{-1}\rho\phi$.  Moreover
\[
\tilde\rho\pi=\delta\sigma,\qquad \delta:=\tilde\rho\rho^{-1}=(i\,j\,k).
\]
\end{lemma}

\begin{proof}
The block move is the permutation
\begin{align*}
\phi(p)&=p &&(p\in A\cup D),\\
\phi(p)&=i+(p-j) &&(p\in C),\\
\phi(p)&=p+(k-j) &&(p\in B),
\end{align*}
which is order preserving inside each block.  Hence
$\tilde\rho(p)=\phi^{-1}(\phi(p)+1)$ equals $p+1\pmod M$ except where the
successor in $z_{0}=A\,C\,B\,D$ leaves a block.  Reading the three junctions
off the new boundary gives (indices in the first column modulo $M$, so
$p=i-1$ is $M-1$ when $i=0$)
\[
\begin{array}{c|c|c}
p & \text{last dart of} & \tilde\rho(p)\\ \hline
i-1 & A\ (\text{or }D\text{ if }i=0) & j\\
j-1 & B & k\\
k-1 & C & i
\end{array}
\]
For $i>0$ the last dart of $D$ still goes to $0$, as under $\rho$; for $i=0$
it is the $i-1$ row of the table and goes to $j$.  Since
$\rho^{-1}(q)=q-1$, evaluating $\delta(q)=\tilde\rho(q-1)$ gives
\[
\begin{array}{c|ccc}
q&i&j&k\\ \hline
q-1&i-1&j-1&k-1\\
\delta(q)&j&k&i
\end{array}
\]
and $\delta(q)=q$ elsewhere, so $\delta=(i\,j\,k)$.  For a concrete check,
$M=10$, $(i,j,k)=(2,5,7)$ has $\tilde\rho(1)=5$, $\tilde\rho(4)=7$,
$\tilde\rho(6)=2$ and $\tilde\rho(p)=p+1$ otherwise, hence
$\delta(2)=5$, $\delta(5)=7$, $\delta(7)=2$.  Finally
$\tilde\rho\pi=(\tilde\rho\rho^{-1})(\rho\pi)=\delta\sigma$.

The transport is a conjugation: $\phi$ carries the letter at $p$ to position
$\phi(p)$, so $\phi\pi\phi^{-1}$ pairs two positions of $z_{0}$ carrying inverse
letters, that is, it is again a pairing.  Its face permutation is conjugate to
$\tilde\rho\pi$, so the two have the same orbit count.

Reading the darts around the single vertex in cyclic order gives the word
$z_{0}=A\,C\,B\,D$ (equivalently, the dual one-face map has boundary $z_{0}$),
which need not be reduced; it spells the element $\mathrm{red}(z_{0})$, and
Culler's theorem bounds $\cl(\mathrm{red}(z_{0}))$ by the genus of the map ---
it refers to the element, not to the spelling --- so nothing asks $z_{0}$ to be
reduced.
\end{proof}

\begin{lemma}[when a $3$-cycle adds two orbits]\label{lem:tau}
Let $\tau=(a\,b\,c)$ and let $\sigma$ be any permutation.  Then
$c(\tau\sigma)-c(\sigma)=2$ if $a,b,c$ lie on one $\sigma$-orbit in the cyclic
order $a\to c\to b\to a$, and $c(\tau\sigma)\le c(\sigma)$ otherwise.
\end{lemma}

\begin{proof}
Write $\tau=(a\,c)(a\,b)$, the right factor acting first.  For a transposition
$t=(x\,y)$ we have $c(t\sigma)=c(\sigma)+1$ if $x$ and $y$ lie on one orbit of
$\sigma$, and $c(\sigma)-1$ if they lie on two.  Set $t_{1}=(a\,b)$ and
$t_{2}=(a\,c)$; we apply these two rules in turn.

If $a,b$ lie on distinct orbits, $t_{1}$ merges them ($-1$) and
$t_{2}$ adds at most $1$, so the net change is non-positive.  Otherwise
$t_{1}$ splits their orbit into the arc
from $a$ to $\sigma^{-1}(b)$ and the arc from $b$ to $\sigma^{-1}(a)$, a gain of
$1$.  If the cyclic order is
$a\to\cdots\to c\to\cdots\to b\to\cdots\to a$, then $c$ lies on the first arc and
$t_{2}$ cuts again, giving $2$.  In the other order, or if $c$ lies on another
orbit, $a$ and $c$ are on distinct orbits of $t_{1}\sigma$ and $t_{2}$ merges
them, giving $0$.
\end{proof}

Call $(i,j,k)$ with $i<j<k$ a \emph{$132$-triple} if its entries lie on one
$\sigma$-orbit in the cyclic order $i\to k\to j\to i$.  With
$\tau=\delta=(i\,j\,k)$, the two lemmas give
\begin{equation}\label{eq:132}
c(\tilde\rho\pi)=c_{0}+2\quad\Longleftrightarrow\quad
(i,j,k)\ \text{is a $132$-triple}.
\end{equation}
So an exact peel exists as soon as $\sigma$ contains a $132$-triple.

\begin{lemma}[a $132$-triple exists]\label{lem:exists}
Let $\pi$ be an optimal pairing of a cyclically reduced $w\in[F,F]$ with
$\cl(w)=\kappa\ge1$.  Then $\sigma=\rho\pi$ has a $132$-triple.
\end{lemma}

\begin{proof}
\emph{Descents.}  Put $d_{p}:=(\sigma(p)-p)\bmod M$ and, for an orbit $C$, let
$D_{C}:=\#\{p\in C:\sigma(p)<p\}$ be its number of descents.  We claim
\[
\sum_{p}d_{p}=M(n+1),\qquad \sum_{C}D_{C}=n+1,\qquad
\sum_{C}(D_{C}-1)=2\kappa\ge2 .
\]
Write each pair of $\pi$ as $\{p,q\}$ with $p<q$.  Since
$\sigma(p)=\pi(p)+1\pmod M$ and $q\ge p+2$,
\[
d_{p}=q+1-p,\qquad d_{q}=M+p+1-q,
\]
so $d_{p}+d_{q}=M+2$, and the $n$ pairs give $\sum_{p}d_{p}=n(M+2)=M(n+1)$
because $M=2n$.  On an orbit $C$ the increments $\sigma(p)-p$ cancel, while
$d_{p}$ exceeds $\sigma(p)-p$ by $M$ exactly when $\sigma(p)<p$; hence
$\sum_{p\in C}d_{p}=M\,D_{C}$.  Comparing the two expressions for
$\sum_{C}\sum_{p\in C}d_{p}$ gives $\sum_{C}D_{C}=n+1$.  Each $D_{C}\ge1$,
because $\sigma(p)>p$ for all $p\in C$ fails at the largest element of $C$.
Since there are $c_{0}=n+1-2\kappa$ orbits,
$\sum_{C}(D_{C}-1)=(n+1)-c_{0}=2\kappa\ge2$.

\emph{Reading a triple off an orbit.}  By the claim some orbit $C$ has
$D_{C}\ge2$; an orbit of two points has exactly one descent, so
$\abs{C}=L\ge3$.  Let $p$ be the least element of $C$ and write its
$\sigma$-cycle as $p=s_{0}\to s_{1}\to\cdots\to s_{L-1}\to p$.  Since $p$ is the
least element and $\sigma$ has no fixed point, $\sigma(p)=s_{1}>p$.  The last
step is a descent, since it returns to the least element, and it is not the only
one,
so the least index $t$ with $s_{t+1}<s_{t}$ satisfies $1\le t\le L-2$.  Then
$p=s_{0}<s_{1}<\cdots<s_{t}$.  Put $i:=p$, $j:=s_{t+1}$ and $k:=s_{t}$.  Then
$i<j<k$, because $s_{t+1}\ne p$ (the cycle meets $p$ only at its two ends), and
the cycle read from $i$ reaches $k$ before $j$, so $(i,j,k)$ is a
$132$-triple.
\end{proof}

\begin{theorem}[exact peel]\label{thm:peel}
Every reduced $w\in[F,F]$ with $\cl(w)=\kappa\ge1$ has cut points
$0\le i\le j\le k\le\abs{w}$ whose residual \eqref{eq:peel} satisfies
$\cl(z)=\kappa-1$; the cut points may be taken with $i<j<k$.
\end{theorem}

\begin{proof}
\emph{The cyclically reduced core.}  Write $w=u\,w'\,u^{-1}$ with $w'$
cyclically reduced (possibly $u=1$), and let $\pi'$ be an optimal pairing of
$w'$; thus $\cl(w')=\cl(w)=\kappa$ and, for
$\sigma'=\rho'\pi'$, $c(\sigma')=n+1-2\kappa$ with $n=\abs{w'}/2$.
Lemma~\ref{lem:exists} applied
to $w'$ and $\pi'$ gives a $132$-triple $(i',j',k')$.  Cut
$w'=A'B'C'D'$ there and set $z'_{0}=A'C'B'D'$.

\emph{The genus drops by one.}  Lemma~\ref{lem:three} and
\eqref{eq:132}, now applied to $w'$, give $c(\tilde\rho'\pi')=c(\sigma')+2$, so
the one-vertex map obtained by transporting $\pi'$ to $z'_{0}$ has genus
\[
\gamma=\tfrac12(n+1-c(\sigma')-2)
=\tfrac12\bigl(n+1-(n+1-2\kappa)-2\bigr)=\kappa-1 ,
\]
and the map has boundary $z'_{0}$ and genus $\gamma$.  A pair $ee^{-1}$ in the
boundary is a spur: in this one-vertex map it bounds a monogon face, so deleting
the edge lowers both the edge count $n$ and the face count $c=c(\tilde\rho'\pi')$
by one, with $V=1$ fixed.  Hence $2\gamma=n+1-c$ and the genus are unchanged.  Repeating removes every spur; the boundary becomes
$\mathrm{red}(z'_{0})$ and the genus stays $\gamma$, so
Proposition~\ref{prop:cycles} gives $\cl(\mathrm{red}(z'_{0}))\le\gamma=\kappa-1$.
Reduction preserves the class, so $\cl(z'_{0})=\cl(\mathrm{red}(z'_{0}))$ ---
Culler's theorem, as noted above, refers to the element, not the spelling.

\emph{Restoring the conjugation.}  Now put $A=uA'$, $B=B'$, $C=C'$,
$D=D'u^{-1}$, so $w=A\,B\,C\,D$, the cut points $i=\abs{u}+i'$, $j=\abs{u}+j'$,
$k=\abs{u}+k'$ are positions of $w$, and $ACBD=u\,z'_{0}\,u^{-1}$.  For the
residual $z=\mathrm{red}(ACBD)$ of \eqref{eq:peel}, reduction and conjugation
preserve the class, so $\cl(z)=\cl(z'_{0})\le\kappa-1$.

\emph{The reverse inequality.}  Conversely, multiplying out as
in Proposition~\ref{prop:peel},
\[
w\,z_{0}^{-1}=A\,B\,C\,D\cdot D^{-1}B^{-1}C^{-1}A^{-1}
=A\,[B,C]\,A^{-1},
\]
so $\kappa=\cl(w)\le1+\cl(z_{0})=1+\cl(z)$ and $\cl(z)\ge\kappa-1$.  The two
bounds agree.
\end{proof}

The algorithm never invokes the theorem: it enumerates the triples of
\eqref{eq:peel} and recurses on the survivors, which is complete by
Lemma~\ref{lem:complete}.

\begin{lemma}[completeness]\label{lem:complete}
Let $r\ne1$ be a reduced residual of even length and trivial abelianisation,
held by a node with target $\mathit{rem}\ge\cl(r)$.  Then the subtree below that
node contains a leaf: a node whose target is $1$, or one whose residual is
empty.
\end{lemma}

\begin{proof}
Induct on $\mathit{rem}$.  For $\mathit{rem}=1$ the node is itself such a leaf:
by Lemma~\ref{lem:onepass} and Proposition~\ref{prop:3hop} the base test
decides $\cl(r)=1$ correctly.  For $\mathit{rem}\ge2$ Theorem~\ref{thm:peel}
applies, because $\cl(r)\ge1$, and gives a peel whose residual $z$ is reduced,
even and of trivial abelianisation, a peel only permuting blocks.  The filters
reject only through \eqref{eq:D}, a lower bound for $\cl(z)$, so this candidate
survives.  If $\cl(r)=1$ then $z=1$ and the child is a leaf with
empty residual; otherwise $\cl(z)=\cl(r)-1\ge1$, so $z\ne1$ and, as
$\mathit{rem}-1\ge\cl(z)$, induction applies to the child.
\end{proof}

\subsection{The filter stack}\label{sec:filters}

A candidate is screened by a single pass over $z$.  The count \eqref{eq:count}
is read on the cyclically reduced core of $z=\mathrm{red}(ACBD)$, so it is
independent of the chosen anchor; it is computed afresh from $z$ itself, never
carried over from the parent, and a cancellation at a junction cannot corrupt
it.  The pass strips the longest mirror-inverse prefix, the largest $x$ with
$z=x\,z'\,x^{-1}$ (the $s_{\max}$ of {\S}2), leaving a cyclically reduced core
$z'$, counts the cyclic
blocks of the core and forms the differences $b_{U}-b_{U^{-1}}$, in
$O(\abs{z})$ time and space, the order of producing $z$ from the four blocks of
the parent.  The candidate is rejected as
soon as $\Dh(z)\ge12(\mathit{rem}-1)-5$: by \eqref{eq:D} this forces
$\cl(z)\ge\mathit{rem}-1+\tfrac1{12}$, hence $\cl(z)\ge\mathit{rem}$, one more
than the child's target, and the rejection rests on \eqref{eq:D} alone.

So a candidate costs $O(M)$ and the filters do not enter the exponent of
Theorem~\ref{thm:rec}, only the constant; their role is practical, discarding a
child whose class provably exceeds the target before any recursion starts.

\subsection{The cost of the recursion}\label{sec:tree}

\emph{The recursion.}  A node holds a reduced residual $r$ of even length and a
target $\mathit{rem}$.  If $\mathit{rem}=1$ it runs the base test of {\S}2;
otherwise it tries every triple $0\le i\le j\le k\le\abs{r}$ of
\eqref{eq:peel}, filters the candidates, and recurses on the survivors with
target $\mathit{rem}-1$; an empty residual ends its branch.

\begin{theorem}[higher-class worst case]\label{thm:rec}
For a word of length $M$ and a target $g\ge2$, the search runs in
$O(M^{3g})$ steps in the unit-cost model, in $O(M^{2})$ space.
\end{theorem}

\begin{proof}
\emph{Correctness.}  For $w=1$ the answer is the empty decomposition, so assume
$w\ne1$.  The
filters never discard a residual whose class is within its target: they reject
only through \eqref{eq:D}.  Hence
Lemma~\ref{lem:complete} applies at every node that is reached, and the search
returns a decomposition into at most $g$ commutators whenever $g\ge\cl(w)$.  For $g<\cl(w)$ it reports
$\cl(w)>g$, which is correct: a leaf at depth at most $g-1$ would have
exhibited $w$ as a product of at most $g$ commutators.  Every returned answer is
a product of commutators equal to $w$: at an internal node by \eqref{eq:exch},
at a base node once the stripped conjugation $r=xzx^{-1}$ is put back.

\emph{The time bound.}  Let $m_{d}\le M$ bound the residual length at depth $d$.  A node with target at
least $2$ enumerates $\binom{m+3}{3}\le(m+1)^{3}$ triples and spends $f(m)=O(m)$
on each, so such a node costs $O(m^{4})$; a node with target $1$ runs the base
test, at $C\,m^{3}$ by Proposition~\ref{prop:base}.  The target, not the
length, drops by one per level, so a node at depth $d$ has target $g-d$ and the
tree has depth at most $g-1$; with $N_{0}=1$ and
$N_{d+1}\le N_{d}(m_{d}+1)^{3}$ this gives $N_{d}\le(M+1)^{3d}$, and
$m_{d}\le M$.  Hence
\begin{align*}
T&\le\sum_{d=0}^{g-2}N_{d}m_{d}^{3}f(m_{d})+N_{g-1}C\,m_{g-1}^{3}\\
 &\le f(M)\sum_{d=0}^{g-2}(M+1)^{3d+3}+C\,(M+1)^{3g}\\
 &=O(M^{3g}),
\end{align*}
when $f(M)=O(M)$.  Here $m_{d}^{3}f(m_{d})=O(m_{d}^{4})$ is the per-node cost:
$m_{d}^{3}$ triples at $f(m_{d})=O(m_{d})$ each.  The sum is dominated by its
last term: the internal layers contribute $O(M^{3g-2})$ against the
$O(M^{3g})$ of the base layer, so the factor $m$ spent on each candidate by
the filter does not raise the exponent.

\emph{The space bound.}  The base test of Proposition~\ref{prop:base} keeps the arc graph,
which has $\nu\le M^{2}$ arcs, while a peel node keeps its residual and the
candidate it is screening, $O(M)$ letters: the enumeration is depth first,
recursing on a survivor as soon as it passes the single-pass filter and
resuming where it left off on return, so no list of survivors is stored.  The stack of residuals holds $O(M)$ letters per
level and the depth is at most $\min(g,\cl(w))\le M$, so it is $O(M^{2})$ as
well, and the total is $O(M^{2})$.
\end{proof}

Nor can two cut points suffice.  A two-cut peel has three blocks and exchanges
the last two, so a two-cut search enumerates $\Theta(M^{2})$ such exchanges per
level and would give $O(M^{2(g-1)+3})=O(M^{2g+1})$.  Ten letters defeat it.  The
word
\begin{equation}\label{eq:ten}
w=w_{0}\cdots w_{9}=a\,b\,a^{-1}c\,b^{-1}d\,c^{-1}e\,d^{-1}e^{-1}
=[ab\,a^{-1},\,c\,a^{-1}]\,[cd\,c^{-1},\,e\,c^{-1}]
\end{equation}
is cyclically reduced of length $10$ and a product of two commutators, but has
no arc, so $\cl(w)=2$ by Proposition~\ref{prop:3hop}.  Its five inverse pairs
sit at $0\!\leftrightarrow\!2$, $1\!\leftrightarrow\!4$,
$3\!\leftrightarrow\!6$, $5\!\leftrightarrow\!8$, $7\!\leftrightarrow\!9$.
A two-cut exchange $ABC\mapsto ACB$ changes only three adjacencies; checking
the $36$ nondegenerate cuts shows that every child has arcs only of length at
most $1$.  Here a child is the residual $\mathrm{red}(ACB)$ of \eqref{eq:peel};
reducing it to its cyclically reduced core costs at most one inverse pair, so
that core still has length at least $8$.  Yet a $c$-word of length at least
$8$ has arcs whose lengths sum to $\mathit{half}\ge4$, hence one of length at
least $2$.  So no child is a $c$-word, no child has class $1$, and a two-cut
search never reaches a leaf.
The exponent is $3g$, not $2g+1$.

\end{document}